# Sub-μ structured Lotus Surfaces Manufacturing

M. Worgull[1], M. Heckele[1], T. Mappes[2], B. Matthis[1], G. Tosello[3], T. Metz[4],
J. Gavillet[5], P. Koltay[4], H. N. Hansen[3]

[1] Forschungszentrum Karlsruhe (FZK), Institute for Microstructure Technology (IMT),
D-76344 Eggenstein-Leopoldshafen, Germany
[2] University of Karlsruhe (TH), Institute for Microstructure Technology (IMT),
D-76344 Eggenstein-Leopoldshafen, Germany
[3] Technical University of Denmark (DTU), Department of Mechanical Engineering (MEK),
DK-2800 Kgs. Lyngby, Denmark
[4] University of Freiburg, Department of Microsystems Engineering (IMTEK),
79110 Freiburg, Germany
[5] French Atomic Energy Commission (CEA), Laboratory of Innovation for New Energy Technologies and
Nanomaterials (LITEN), 38054 Grenoble, France

**Abstract**

Sub-micro structured surfaces allow modifying the behavior of polymer films or components. Especially in micro fluidics a lotus-like characteristic is requested for many applications. Structure details with a high aspect ratio are necessary to decouple the bottom and the top of the functional layer. Unlike to stochastic methods, patterning with a LIGA-mold insert it is possible to structure surfaces very uniformly or even with controlled variations (e.g. with gradients). In this paper we present the process chain to realize polymer sub-micro structures with minimum lateral feature size of 400 nm and up to 4 μm high.

I. INTRODUCTION

A new generation of passive, capillary driven micro-fluidics systems is expected to enable advanced management of liquid and gas. In such enhanced flow structures, fluids are guided through channels with geometrically modified surfaces. In particular, surface sub-μ structuring of micro fluidic systems can be designed to introduce a structural gradient which can generate a driving force to move liquid samples along channel structures (see Figure 1). The realization of Lotus–like patterns on surfaces requires precise control of the structuring method. The manufacturing method of such structured surfaces should also enable mass-fabrication capability, for example by polymer replication. To demonstrate the moulding of sub-μ structures, an indirect tooling technology for the manufacturing of the mould insert and the hot embossing process for polymer replication has been selected. In this paper we present the process chain to realize polymer sub-micro structures with minimum lateral feature size of 400 nm and up to 4 μm high.

2. WETTING BEHAVIOR OF STRUCTURED SURFACES

The liquid repellency of a surface is principally governed by a combination of its chemical nature (i.e. surface energy) and, in case of stochastic surfaces, by its topography at the micro-scale (i.e. surface roughness). Although flat low surface energy materials can often exhibit high water contact angles [1] this is normally not sufficient to yield the reppellency of super-hydrophobic surfaces. In order to obtain this, the difference between the advancing and the receding contact angle (contact angle hysteresis) must be minimal. Effectively, contact angle hysteresis is the force required to move a liquid droplet across a surface. In the case of little or no hysteresis, very little force is required to move a droplet, hence it rolls off easily [2][3].

Theoretical studies for idealized rough hydrophobic surfaces predict that contact angle hysteresis initially increases with surface roughness [4] until eventually a maximum value is reached. Greater roughness scales beyond this lead to a decline of the contact angle hysteresis due to the formation of a composite interface (where the liquid is unable to completely penetrate the surface).

The phenomena can be described by the Cassie-Baxter equation (1), where porous surface topography, superimposed to the roughness of a solid surface, causes air to become trapped in voids (i.e. prevents liquid from wicking) [5]:

$$\cos \vartheta^* = f \cdot \cos \vartheta + f - 1 \qquad (1)$$

Where: θ* is the apparent contact angle, f is the surface fraction (i.e. the total area of solid-liquid interface in a unity of plane geometrical area parallel to the surface), and θ is the contact angle of the rough surface but not porous.

The surface fraction for a simple porous surface structure such as a columnar structure with square section is defined as follows (see Figure 1):

$$f = \frac{a^2}{(a+b)^2} \quad (2)$$

Where: *a* is the pillar width and *b* is the spacing between two consecutive pillars (see Figure 1). By increasing the void between two micro structures (i.e. decreasing the width of the structures), a smaller surface fraction can be obtained (see Equation (2)) and therefore a larger contact angle (see Equation (1)) is achieved. A hydrophobic substrate can become super-hydrophobic; under given circumstances even a hydrophilic substrate can act as a hydrophobic one.

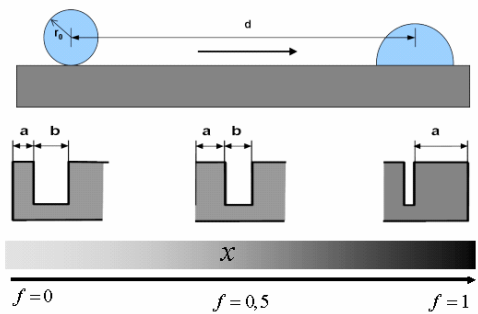

Figure 1 – Passive droplet activation by means of a linear gradient, i.e. a change of the surface fraction *f* along the moving direction obtained by different openings *b* on the surface's substrate.

### 3. DESIGN OF LOTUS STRUCTURES

In this work honeycomb structures are used to and can be attained over a broad range of geometric parameters. To compare different wetting behaviors depending on the ratio between void and bulk characteristic lengths (see Figure 1), two different designs were investigated and manufactured (see Figure 2). The pitch of the micro-structured pattern was maintained constant (pitch = 4 μm), whereas the diameter of the combs and the thickness of the separating walls varied. The two patterns had the following geometrical characteristics:

1. Wall thickness = 1000 nm and comb diameter = 3000 nm (opening ratio f=0.25).
2. Wall thickness = 400 nm and comb diameter = 3600 nm (opening ratio f=0.10).

The height of the structures was 4 μm resulting on an aspect ratio of the sub-micro walls between 4 and 10. The height of the structures (and therefore the aspect ratio) was chosen in order to obtain superhydrophobic surfaces and also to allow the demolding after the replication by hot embossing, without damaging the separating walls.

The structured surface was of an area of 20x10mm² split into two different areas of 10x10mm² with a different diameter and pitch of the honeycomb structures. The two designs were placed as close as possible to each other to analyze an effect at the interface between both designs (see Figure 2).

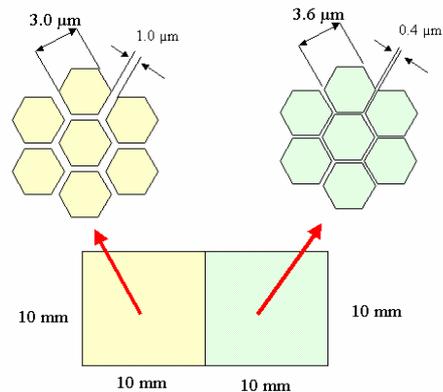

Figure 2: Design and arrangement of honeycomb structures

### 4. PROCESS CHAIN

The process chain consist of the process steps E-beam lithography, mask fabrication, X-ray lithography, electroplating of mold insert and finally replication by hot embossing. (Figure 3)

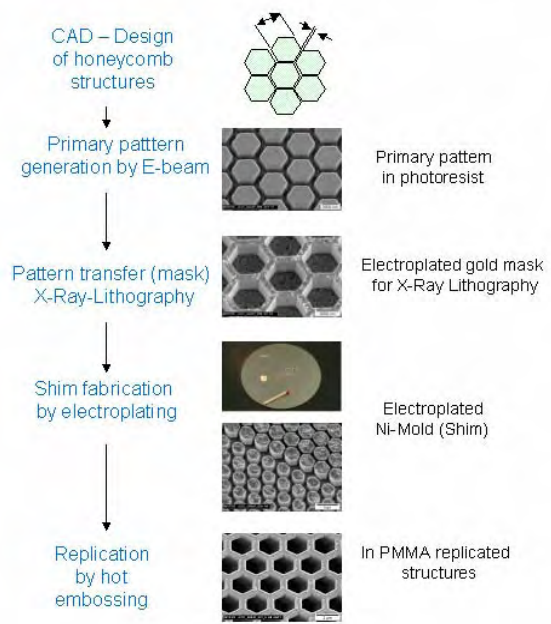

Figure 3 - Process chain for fabrication of sub-micron Lotus structures by E-beam lithography, mask fabrication, X-Ray lithography, electro forming and finally replication by hot embossing.

### 4.1 E-BEAM LITHOGRAPHY

The design was written via electron beam (E-beam) lithography into a 3.2 µm thick photoresist (PMMA). The E-beam device was operated at maximum acceleration voltage of 100 kV. In order to avoid damage of structures during the demolding step of the hot embossing process, vertical sidewalls on the mold insert have to be manufactured. Hence, to avoid any undercuts in the mold inserts already at the beginning of the process chain, the dose during E-beam lithography had to be optimized. Therefore variations in dose had to be performed to calibrate the E-beam process and to select the beam parameters for the writing procedure. Finally, writing was performed with a current of 10nA and a dose of 800µC/cm$^2$ (see Figure 4a, 4b).

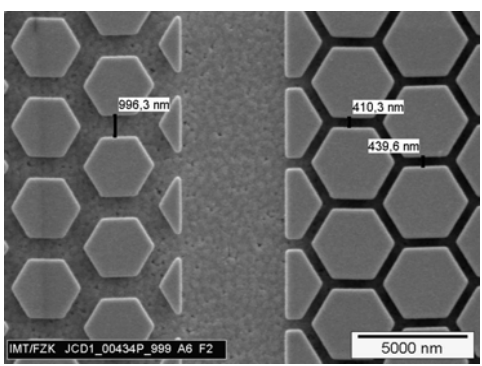

Figure 4a: Intersection of the two different honeycomb designs after E-beam lithography

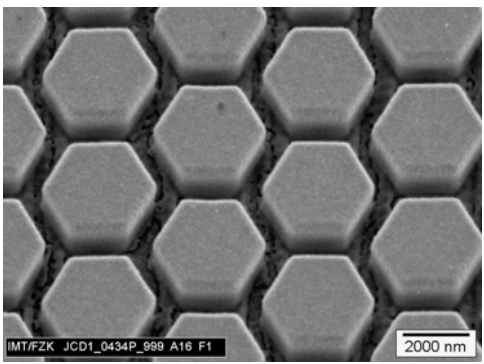

Figure 4b. Detailed view of the honeycomb structures after E-beam lithography. The thickness of the resist is 3 µm.

### 4.2 MASK FABRICATION

The generated polymer grooveswere filled with gold by electroplating (see Figure 5) and developed. To obtain a gold layer with a homogeneous absorption a gold sulfidic bath was employed. The thickness of the gold mask was in the range of 2.20 ± 0.05 µm.

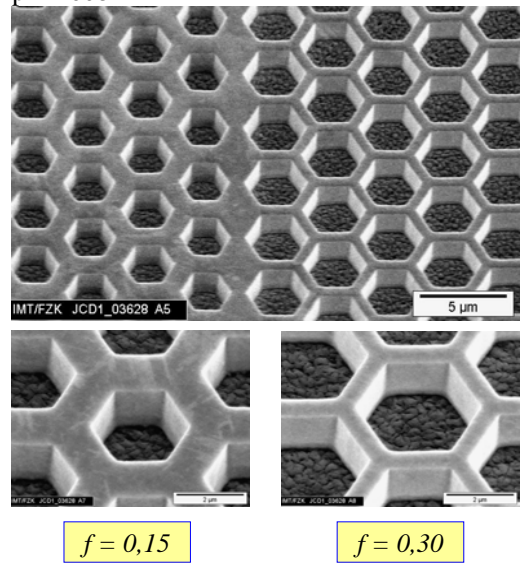

*f = 0,15*   *f = 0,30*

Figure 5 – Electroplated gold mask containing honeycomb sub-µ structures with two different aspect ratios. Height of the structures is 2.5 µm. A different surface fraction is obtained by decreasing the wall thickness from 1200 nm (aspect ratio = 2.1) to 600 nm (aspect ratio = 4.2) (left and right respectively) with a unit cell feature pitch of 4 µm.

### 4.3 X-RAY LITHOGRAPHY

The mask was used in a next step for the transfer of the design to a 4 µm and a 10 µm thick resist of PMMA. Here different aspect ratios of the structures up to 20 (referring to the sidewalls) were achieved. As shown in Figure 6, a deformation of the structures of 10 µm thickness occured due to a shift of the mask during the X-ray lithography process.

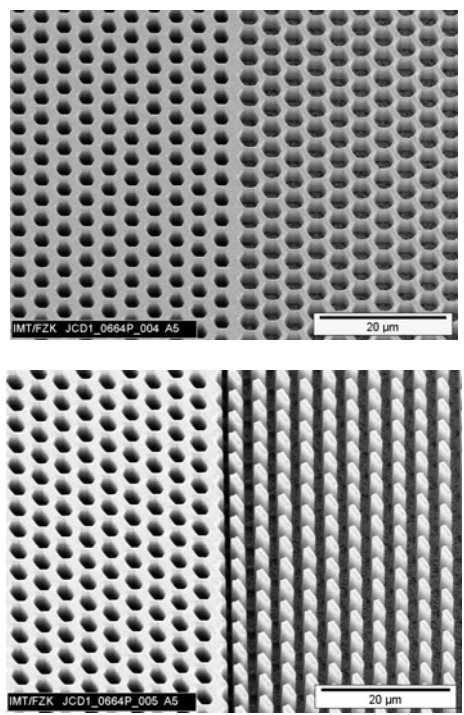

Figure 6. Via X-Ray lithography the structures were transferred into 4 µm high resist (top picture) and 10 µm high resist (bottom picture).

### 4.4 ELECTROPLATING OF NI-SHIM

Based on the structures fabricated by X-ray lithography two nickel mold inserts were fabricated by electroplating. The mold inserts were later used for replication by hot embossing. The two fabricated shims had a diameter of 4 inch and a thickness of approximately 300 µm. Representative for these mold inserts, a nickel shim with optical structures and similar dimensions of structures, is shown in Figure 7. Honeycomb structures of the microstructured nickel mold insert with 4µm height areshown in Figure 8.

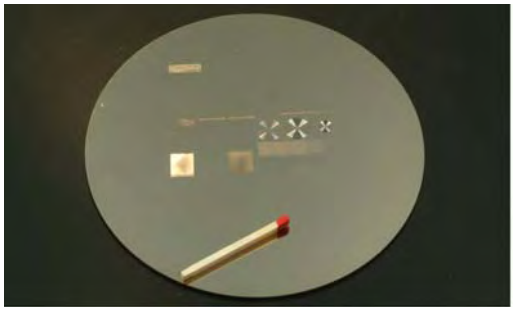

Figure 7 Electroplated microstructured Nickel shim. Nickel shims with a typical thickness of about 300 µm are well suited for the replication of nanostructures . Typical diameters are in the range of 4 inch or 6 inch.

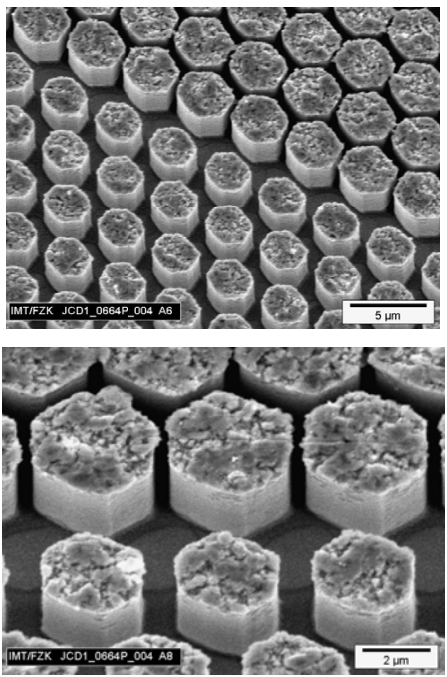

Figure 8. Honeycomb structures of the microstructured nickel mold insert. In the pictures the electroplated shim from the 4 µm thick resist is shown. The detail view on the bottom picture shows the rough surface on the top side of the structures as an result of the roughness of the adhesion and starting layer TiOx.

### 4.5 REPLICATION BY HOT EMBOSSING

The replication of the mold inserts into PMMA was carried out by hot embossing (Figure 9). Hot embossing is well suited for the replication of large structured areas with high aspect ratio sub-µ structures on a thin residual layer [6]. Especially the short flow distances and the moderate flow velocities produce low stress in the molded part. Furthermore, the precise vertical demolding, guaranteed by the precise vertical guidance of the crossbars of the machine, is essential for successful demolding of the filigree lotus structures. Honeycomb structures in the submicron range with an aspect ratio of 3.5 were already replicated successfully by this replication technology [7]. Nevertheless, because of the influence of shrinkage of the polymer, a challenging task is the demolding without deformation or damage of the structures. This holds in particular for the honeycomb structures with sidewalls in the range of 400 nm and structure height in the range of 10 µm.

The replication of high aspect ratio micro-nano structures into polymer with a very low defect rate is essential for the functionality of the superhydrophobic surface. The lotus structures can be replicated in a wide range of polymer materials, beginning from low temperature materials like PS or PMMA up to semi-crystalline high temperature materials like LCP or PEEK. For the first approach PMMA was employed. An example of sub-micron honeycomb structures molded in PMMA by hot embossing is shown in Figure 10. The structure size of these honeycombs is in the same range like as structures fabricated in this work.

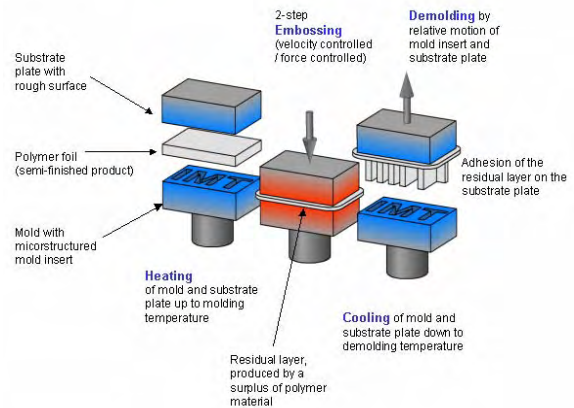

Figure 9 Schematic view of the hot embossing process. The process is characterized by heating of a polymer film via heat conduction, a two step embossing cycle, a convective cooling and finally a vertical demolding.

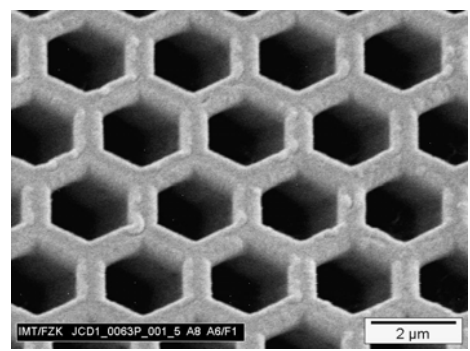

Fig. 10. Sub-µm honeycomb structures replicated in PMMA by hot embossing (pattern pitch is 2 µm and wall thickness is 500 nm.

## 5. FUNCTIONAL TEST

The polymer (PMMA) substrate structured via hot embossing was tested by means of contact angle measurements. Droplets of distilled water where deposited on the three different areas of the sample (droplet volume 1.1±0.1 µl, see Figure 10). Measured contact angle were the following:
- Reference area (without structuring) = 81±4 °
- Area 1 (1000 nm thick walls) = 87±2 °
- Area 2 (400 nm thick walls) = 107±6 °

By introducing the sub-µm honeycomb structures on the surface, the hydrophilic property of the PMMA substrate (contact angle lower than 90 °) has been turned apparently into a hydrophobic one. A decrease of the wall thickness has produced an increase in contact angle (see Figure 11).

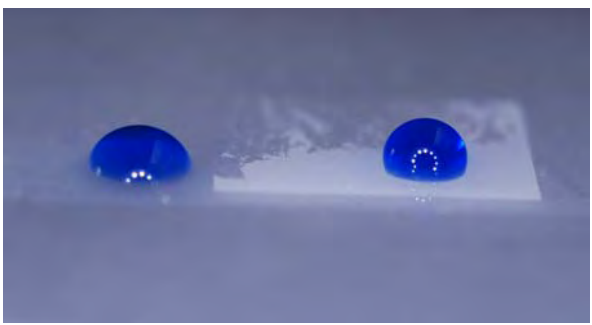

Figure 11. Different wetting behaviors due to the surface sub-micro structuring of the polymer substrate: honeycomb with wall thickness of 400nm (right), honeycomb with wall thickness of 1000nm (left).

## 6. SUMMARY

A new process chain for the replication of sub-µm structured surfaces into polymer was established. The tooling process includes E-beam writing, mask fabrication and X-ray exposure with subsequent nickel electroplating. Hot embossing was used to replicate on a polymer substrate the micro-nano honeycomb structures.

Surface sub-µm structuring can be employed to provide particular properties to the polymer surface for enhanced flow functionality suitable for microfluidics applications. In this research work, in particular, sub-µm honeycomb structures were used to provide hydrophobic properties to a hydrophilic polymer surface. Alternatively, hydrophobic substrates can be turned into super- hydrophobic by using the presented surface structuring.

Advanced design solutions employing this new technology may include micro fluidic systems whose surfaces incorporates structural gradients generating a driving force to move liquid samples along channel structures. Figure 12 shows a gradient design which will be proceed with the same process chain described above.

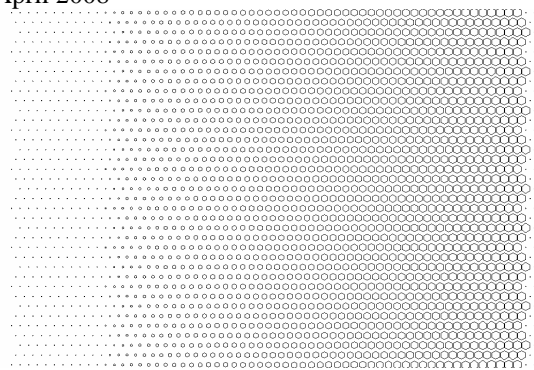

Figure 12: Lotus gradient design generating a driving force to move liquid samples along channel structures

## 7. ACKNOWLEDGEMENTS

The presented research was performed within the framework of the European Network of Excellence "Multi Material Micro Manufacture: Technology and Applications" (4M) (European Community founding FP6-500274-1; www.4m-net.org) and within the activity of the Cross Divisional Project "MINAFLOT" (Micro- and Nano-structured Surfaces for Liquid and Gas Management in Microstructured Flowfields) supported by 4M and the European Community (Project no. FP6-500274-2). Furthermore the collaboration of the Polymer Technology Division (4M Work package 4) and of the Microfluidics Application Division (4M Work package 10) is gratefully acknowledged.